\documentclass[aip,pof,numerical]{revtex4-1}

\usepackage{dcolumn}%
\usepackage{bm}%
\usepackage[latin1]{inputenc}
\usepackage[english]{babel}
\usepackage[T1]{fontenc}

\usepackage{textcomp}
\usepackage{graphicx}
\usepackage{color}

\begin{document}

\title{Melting dynamics of large ice balls in a turbulent swirling flow}

\author{N.~Machicoane}
\affiliation{Laboratoire de Physique, ENS de Lyon, UMR CNRS 5672, Universit\'e de Lyon, France}
\email{nathanael.machicoane@ens-lyon.org}
\author{J.~Bonaventure}
\affiliation{Laboratoire de Physique, ENS de Lyon, UMR CNRS 5672, Universit\'e de Lyon, France}
\author{R.~Volk}
\affiliation{Laboratoire de Physique, ENS de Lyon, UMR CNRS 5672, Universit\'e de Lyon, France}

\begin{abstract}
We study the melting dynamics of large ice balls in a turbulent von K\'arm\'an flow at very high Reynolds number. Using an optical shadowgraphy setup, we record the time evolution of particle sizes. We study the heat transfer as a function of the particle scale Reynolds number $\mbox{\textit{Re}}_{D}$ for three cases: fixed ice balls melting in a region of strong turbulence with zero mean flow, fixed ice balls melting under the action of a strong mean flow with lower fluctuations, and ice balls freely advected in the whole flow. For the fixed particles cases, heat transfer is observed to be much stronger than in laminar flows, the Nusselt number behaving as a power law  of the Reynolds number\string: $\mbox{\textit{Nu}} \propto \mbox{\textit{Re}}_{D}^{0.8}$. For freely advected ice balls, the turbulent transfer is further enhanced and the Nusselt number is proportional to the Reynolds number $\mbox{\textit{Nu}} \propto \mbox{\textit{Re}}_{D}$. The surface heat flux is then independent of the particles size, leading to an ultimate regime of heat transfer reached when the thermal boundary layer is fully turbulent.\end{abstract}

\date{\today}
\maketitle

\section{Introduction}
Mass or heat transfer from a particle transported by a turbulent flow is encountered in many natural or industrial processes, such as solid dissolution in liquids, droplets vaporization in engines, or ice particles melting in heat exchangers. This problem is complex because it depends on the relative motion between the particle and the fluid, a function not only of the properties of the flow, both also of the particles characteristics. Indeed, material particles with a density differing from that of the fluid, or with a diameter $D$ larger than the Kolmogorov scale $\eta$ are known not to behave as tracers of the flow motions \cite{calzavarini:2008a,bib:qureshi2008_EPJB,volk:2011}. Every transported particle will then explore the flow differently and will dissolve or melt at a different rate depending on its trajectory.\\

Since the applications are of great interest, many experimental studies of heat transfer have been conducted. For instance, concerning small particles, mass transfer was investigated for many particles dissolving in mixers by measuring the evolution of global quantities (conductivity, absorbance, ...) in the first steps of the process \cite{Sano1974,Boon-Long1978}. Separately, evaporation of single droplets, maintained fixed, were performed by recording the evolution of their radius in zero mean turbulent flows \cite{Birouk2002}. All of these studies show that the Nusselt $\mbox{\textit{Nu}}$ (representing heat transfer; or Sherwood number $\mbox{\textit{Sh}}$ representing mass transfer) is a function of the particle Reynolds number $\mbox{\textit{Re}}_{D}$ and Prantl number $\mbox{\textit{Pr}}$. A classical example is the Ranz-Marshall correlation  for heat transfer from a fixed sphere of diameter $D$ undergoing a uniform mean velocity field $U$. When the Reynolds number $\mbox{\textit{Re}}_{D}=UD/\nu$ is small enough, $\nu$ being the fluid kinematic viscosity, the Nusselt number empirically follows \cite{bib:ranz1952}:

\begin{equation}
\mbox{\textit{Nu}} = 2+0.6\sqrt{\mbox{\textit{Re}}_{D}}\mbox{\textit{Pr}}^{1/3}.
\label{eq:ranz}
\end{equation}

For particles larger than the Kolmogorov size $\eta$, the Nusselt number $\mbox{\textit{Nu}}$ is still generally expressed as a power law of the Reynolds number $\mbox{\textit{Re}}_{D}$ \cite{Levins1972,Birouk2002} with an exponent increasing with $\mbox{\textit{Re}}_{D}$ (see Birouk \textit{et al}.\cite{bib:birouk2006} and references therein).\\

Heat transfer from large objects maintained fixed, with sizes of the order of the integral scale, was investigated both numerically and experimentally in turbulent flows where both mean velocity and turbulent intensity can be changed separately \cite{Bagchi2008,Bogusawski2007,vander}. Although simulations seem to indicate a weak impact of the turbulence level on the mean heat transfer \cite{Bagchi2008}, experiments conducted with large heated cylinders or spheres concluded turbulence always increases the heat flux at the particle surface \cite{vander,Bogusawski2007}. This increase was also observed for smaller objects, and a large variety of correlations accounting for the separate influence of mean velocity and turbulence level has been proposed \cite{bib:birouk2006}.\\

For objects whose diameters $D$ are of the order of the integral length scale of the flow, no study about the heat or mass transfer between freely advected particles and the driving turbulent flow has been conducted. Besides, Lagrangian studies of this problem are only a recent matter \cite{bib:zimmermann2011_PRL,Klein2013} because it requires to track the particles along their trajectories for long times while measuring their angular velocity.\\ The present study aims at investigating heat transfer from such large spherical particles when freely advected by a fully turbulent flow. In the case of such large objects, the time averaged sliding velocity seen by the particle is unknown. However one may expect it to be in between the extreme cases of fixed particles suspended either in a zero mean turbulent flow, or in a turbulent flow with mean velocity much larger than fluctuations. We thus investigate the melting of freely advected spherical ice balls in a turbulent flow of water, and contrast our results to situations for which the ice balls are maintained fixed, and submitted to zero mean turbulence, or to turbulent fluctuations in the presence of a mean velocity.
To measure heat transfer, we coupled shadowgraphy and particle tracking to measure the size evolution of every single ice balls along their trajectories while melting.  
Such simultaneous measurements of both size and position of objects in turbulence was proven effective with a holography-based setup in the case of small evaporating Freon droplets in a zero mean turbulent flow \cite{Chareyron2012}.\\
In the following, section II is devoted to the experimental setup description, with the flow configurations and the making of the ice balls used in the study. We then describe the shadowgraphy setup in section III together with the image analysis and calibration of the heat flux measurement. We then present results obtained for the three configurations in Section IV, where we show that freely advected particles melt in the ultimate regime of heat transfer for which the Nusselt number is proportional to the particle Reynolds number. The last section is then devoted to discussion and conclusion of the results.

\section{Experimental setup}
\subsection{The von K\'arm\'an flow}
The experimental apparatus is a von K\'arm\'an flow, similar to the one used in \cite{volk:2011}, the main difference being the shape of the tank, of square section rather than circular, for better optical access. The flow is produced by two discs, fitted with $8$ straight blades, rotating at constant frequency $\Omega$ to impose an inertial steering. The discs radius is $R=7,\!1$ cm and they are spaced by $15$ cm, which is also the length of the tank section. The rotation axis, noted $\hat{z}$, is perpendicular to gravity $\vec g = -g \hat{y}$. The top wall of the tank has a centred hole on which is mounted a tube ($10$ cm in length, $5$ cm in diameter) used to insert large particles into the flow. We use distilled water as a working fluid and a water circulation in the shafts of the vessel behind the discs in order to impose a constant water temperature using a thermal bath. Together with a cooling of the room, this allows for thermalization of the flow between 3 and $20\pm0.1$°C even at the highest Reynolds numbers. Before doing any experiments, we wait for thermal equilibrium between warming from the DC motors or mechanical power injected by the discs into the flow, and the cooling from the thermal bath. We then precisely measure the equilibrium water temperature using a resistance thermometer (Pt100 sensor).

\begin{figure}[h]
  \begin{center}
    \includegraphics[width=\columnwidth]{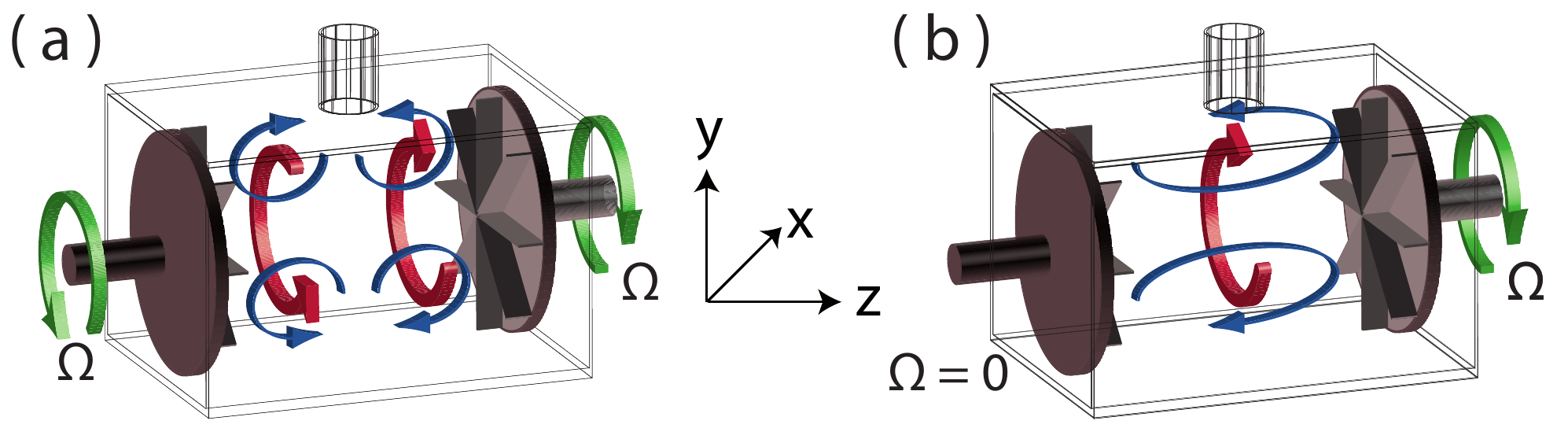}
    \caption{(colour online) Drawing of the mean flow in a square section von K\'{a}rm\'{a}n apparatus. (a): two counter-rotating discs product two azimuthal counter-rotating cells (red arrows) and meridional recirculations (blue arrows). (b): one disc products a strong azimuthal rotation (red arrow) and a meridional recirculation (blue arrows).}
    \label{mean_flow}
  \end{center}
\end{figure}

We use two different flow configurations. On the first hand, when both discs are rotating at same frequency $\Omega$ but with opposite direction (figure \ref{mean_flow}(a)), the mean flow is composed of two counter-rotating cells with azimuthal motion and two meridional recirculations. The two discs configuration produces very intense turbulence: near the geometrical centre, where the mean flow vanishes, fluctuating velocities are of the order of $u'/U \sim 20\%$, $U=2 \pi R \Omega$ being the discs velocity and $u'=\sqrt{({u'_x}^2+{u'_y}^2+{u'_z}^2)/3}$ the magnitude of the velocity fluctuation components $u'_{i}$. At this location, turbulence is nearly homogeneous,  but not isotropic, with fluctuating transverse velocity components ($u'_x$,$u'_y$) 1.5 times the axial component $u'_z$ (see table \ref{tab:flow_2d} for more details). Away from the centre, the mean flow is stronger with less intense turbulent fluctuations \cite{zocchi:1994}. This possibility of having a vanishing mean flow at the centre, or strong mean flow near the discs made this flow very common for studies of turbulence in both Eulerian framework \cite{douady:1991,zocchi:1994} or Lagrangian framework \cite{laporta:2001,mordant:2001,Voth:2002}.\\
On the other hand, when only one disc is rotating, the mean flow is composed of a strong rotating azimuthal motion and a meridional recirculation. Near the geometrical centre the turbulence is homogeneous and isotropic with a strong mean velocity $\langle u_z \rangle$ aligned with the rotation axis, the turbulence level being $u'_z/\langle u_z \rangle \sim 35\%$, which corresponds to $u'_z/U \sim 10\%$ (see table \ref{tab:flow_1d} for more details).\\
Thus the two flow configurations have similar large scale Reynolds number $\mbox{\textit{Re}}=UR/\nu \sim 10^4-10^5$, with different mean flow geometries and turbulence intensities. The one disc configuration leads to fully developed turbulence for the whole range of rotation frequencies under study, all velocities being proportional to $U=2 \pi R \Omega$, which is only the case for $\Omega \geq 4$Hz for the 2 discs flow.

\begin{table*}
\begin{center}
\begin{tabular}{*{7}{cp{.1	cm}}}
   $\Omega$ (Hz) && $u'_z$ (m.s$^{-1}$) && $u'_{(x,y)}$ (m.s$^{-1}$) && $u'$ (m.s$^{-1}$) && $\mbox{\textit{Re}}_{D}$  \\
   \hline
    1.5 && 0.09 && 0.14 && 0.13 && [5,~ 15]$\cdot 10^3$ \\
    4.4 && 0.29 && 0.47 && 0.42 && [15,~ 45]$\cdot 10^3$  \\
    7.3 && 0.48 && 0.76 && 0.68 && [25,~ 75]$\cdot 10^3$\\
\end{tabular}
\end{center}
\caption{Parameters for the two discs flow at several rotating frequencies $\Omega$. $u'_{z}$ and $u'_{(x,y)}$ are the root mean square of the velocity fluctuations in the axial and transverse directions. Their mean $u'=\sqrt{({u'_x}^2+{u'_y}^2+{u'_z}^2)/3}$ is used as the magnitude of the velocity fluctuations. $\mbox{\textit{Re}}_{D}$ is the Reynolds number based on the ice balls diameters ($D=$ [10 - 30] mm) and the discs velocity $U=2 \pi R \Omega$: $\mbox{\textit{Re}}_{D}=UD/\nu$, with $\nu$ being the kinematic viscosity (taken for water at 10°C for tables \ref{tab:flow_2d} and \ref{tab:flow_1d}). Velocity were measured by doppler velocimetry using a PDI from Artium Technologies and 10 $\mu$m tracer particles.}
\label{tab:flow_2d}
\end{table*}

\begin{table*}
\begin{center}
\begin{tabular}{*{7}{cp{.1	cm}}}
   $\Omega$ (Hz) && $\left\langle u_{z} \right\rangle $ (m.s$^{-1}$) && $u'_{z}$ (m.s$^{-1}$) && $u_{trms}$ (m.s$^{-1}$) && $\mbox{\textit{Re}}_{D}$ \\
   \hline
    1.5 && 0.13 && 0.06 && 0.14 && [5,~15]$\cdot 10^3$  \\
    4.4 && 0.42 && 0.15 && 0.44 && [15,~45]$\cdot 10^3$  \\
    7.3 && 0.74 && 0.30 && 0.79 && [25,~75]$\cdot 10^3$  \\
\end{tabular}
\end{center}
\caption{Parameters for the one disc flow at several rotating frequencies $\Omega$. $\left\langle u_{z} \right\rangle $ and $u'_{z}$ are the mean velocity and the root mean square of the velocity fluctuations in the axial direction. $u_{trms}=\sqrt{\left\langle u_{z} \right\rangle^2 + (u'_z)^2}$ is the true rms value of the velocity in the $z$ direction. See table \ref{tab:flow_2d} for other parameters definitions.}
\label{tab:flow_1d}
\end{table*}

\subsection{Making of spherical ice balls}
The ice balls used in the experiments are designed using moulds with spherical prints of diameters $10$, $14$, $18$, $24$ and $30$ mm. The ice balls sizes are of the order  of the discs radius which corresponds to the size of the largest eddies of the flow. They are thus much larger than the Kolmogorov micro-scale $\eta$ (of the order of $20\mu$m \cite{bib:volk2007_EPL}) and do not follow the small scale motions when freely advected by the flow. After the ice balls are made, they are thermalized at their melting temperature $0^\circ$C, so that no diffusion inside the ice balls happens during their melting. Once thermalized, the ice balls can be used for the experiments where they would melt in a flow at a fixed temperature $T_{water}$. Two cases are studied: either ice balls are maintained fixed at the geometrical centre of the flow by a 2 mm PEEK-made rod, or ice balls are freely advected by the flow. PEEK was chosen because of its good mechanical resistance and good thermal insulating properties. For freely advected particles, only the $2$ discs flow is used. For fixed ice balls, both flow configurations are used to understand the influence of fluctuations and time averaged sliding velocity on heat transfer. Indeed, both flows have fluctuating velocity of the same order of magnitude, but the one disc flow has a strong mean velocity at the particle position (large sliding velocity) whereas there is no mean velocity in the centre for the two discs flow (no sliding velocity in average).

\section{Measurement setup}
\subsection{Optical setup}

\begin{figure*}[t]
\centering
\includegraphics[width=1.0\textwidth]{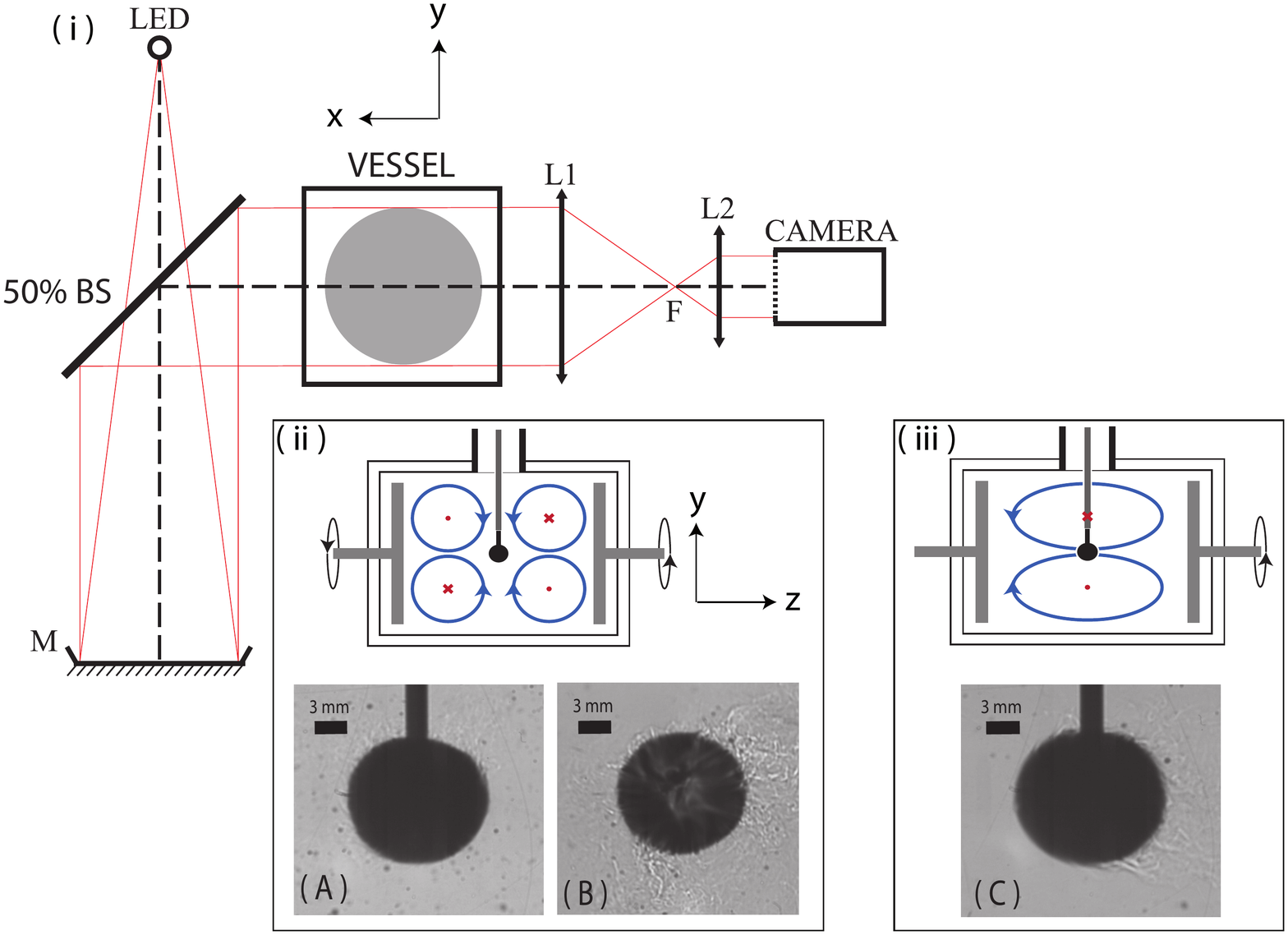}    
\caption{(colour online) (i) : drawing of the optical setup. (ii) : schematic of the 2 discs flow configuration with position of the particle when maintained fixed and corresponding raw image A as obtained with the camera. Image B shows corresponding close up image for a freely advected ice ball melting in the flow. (iii) : schematic of the 1 discs configuration with position of the particle when maintained fixed and corresponding raw image C. The use of parallel light ensures the possibility of defining the particle boundary even in the case of small optical mismatch, with less than $1.5\%$ variations in apparent diameter. Ice balls would be invisible with classical back lighting as usually used for particle size measurements \cite{Birouk2002}.}
  \label{img}
\end{figure*}

The optical setup is designed to measure the size of moving particles in a large portion of space with one camera. To perform this measurement, common optical arrangement cannot be used because the apparent size of the particle changes with the camera to particle distance. We then use an afocal shadowgraphy setup with parallel lighting (figure \ref{img}(i)) for which the apparent size of the particle is independent of its position. A small LED is positioned in the focus of a large parabolic mirror ($15$ cm diameter, $50$ cm focal length) that transforms the diverging light ray emitting from the LED through a $50\%$ beam splitter into a parallel light ray of $15$ cm diameter. This ray of light reflects on the beam splitter and intersects approximately  $80\%$ of the flow volume. It is then collected onto a Phantom V.10 camera (4Mpix@400Hz) whose objective is replaced by a telescope made of two lenses of diameters $15$ cm and $5$ cm with focal lengths $50$ cm and $10$ cm respectively.

With this shadowgraphy optical configuration, particles appear as black shadows on a white background (figure \ref{img}(A-C)), it is then possible to measure their size and shape as they are maintained fixed by the rod or freely advected in the measurement volume. Besides, this setup allows for the sizing of particles with optical index of refraction close to the one of the liquid because the intensity is related to the second derivative of the optical index. For ice particles melting in water it is then possible to define a boundary between solid and liquid phases on the pictures even in the presence of thermal fluctuations in the vicinity of the boundary (figure \ref{img}(A-C)). For all experiment, the ice ball always fill more than 500 pixels in area on the pictures, which allows a good accuracy for the radius detection.

Since the apparent particle size varies only weakly with its position, calibration is straightforward and is performed with moving spherical particles of known sizes for which we can estimate the size measurement error. In the range of ice balls diameters used, we found that the variation of the particle apparent size only changed by less than $1.5\%$ of the true radius. This error accounts for the very small deviation of the ray of light to parallelism, as the LED is not truly point-like. The measurement accuracy could be increased with a stereoscopic setup using an additional camera as it would allow for a 3D calibration using Tsai camera model \cite{bib:tsai1987}. We found it unnecessary for the present study as the bias induced by varying the particle position is much smaller than the particle radii. For both fixed particle cases, the accuracy is even better, because the particle is not moving.


\subsection{Heat flux measurement}
From a sequence of images it is possible to estimate the mean heat flux per unit area, noted $Q_{S}$, at the surface of the melting particle. For particles with imposed temperature $T_S$ at the boundary and when forced convection is overwhelming natural convection, the convective heat flux is proportional to the temperature difference between the water temperature (noted $T_{water}$) and $T_S$. We write $Q_{S}=h (T_{water}-T_S)$, $h$ being the heat transfer coefficient accounting for forced convection. For a particle melting close to equilibrium, one expects $T_S$ to be the melting temperature $T_0$ so that the evolution of the particle volume $V$ (and surface $S$) are governed by Stefan's equation \cite{tranfert_thermique}:

\begin{equation}
\rho_{p}L_{f} \frac{dV}{dt} = \lambda_{th}  S \left \langle \frac{\partial T}{\partial n} \right \rangle_S -  h(T_{water}-T_{0})S.
\label{eq:stefan}
\end{equation}

In this equation $\lambda_{th}$, $\rho_{p}$, and $L_{f}$ are respectively the thermal conductivity, density, and fusion enthalpy of ice at $T=0^\circ$ C, and $\left \langle \frac{\displaystyle  \partial T}{\displaystyle \partial n} \right \rangle_S$ is the normal temperature gradient inside the particle averaged over its surface $S$. A simplification of equation (\ref{eq:stefan}) is obtained in the case of particles initially thermalized at $T=T_0$, for which the diffusion term (due to the temperature gradient inside the particle) disappears. For such a case, one expects the melting rate of the particle $S^{-1}dV/dt$ to be proportional to the temperature difference at a given flow regime. As the melting rate reduces to the derivative of the radius for a sphere, we call it melting speed and note it $dR/dt$ in the following.

\begin{figure}[h]
  \begin{center}
    \includegraphics[width=.49\columnwidth]{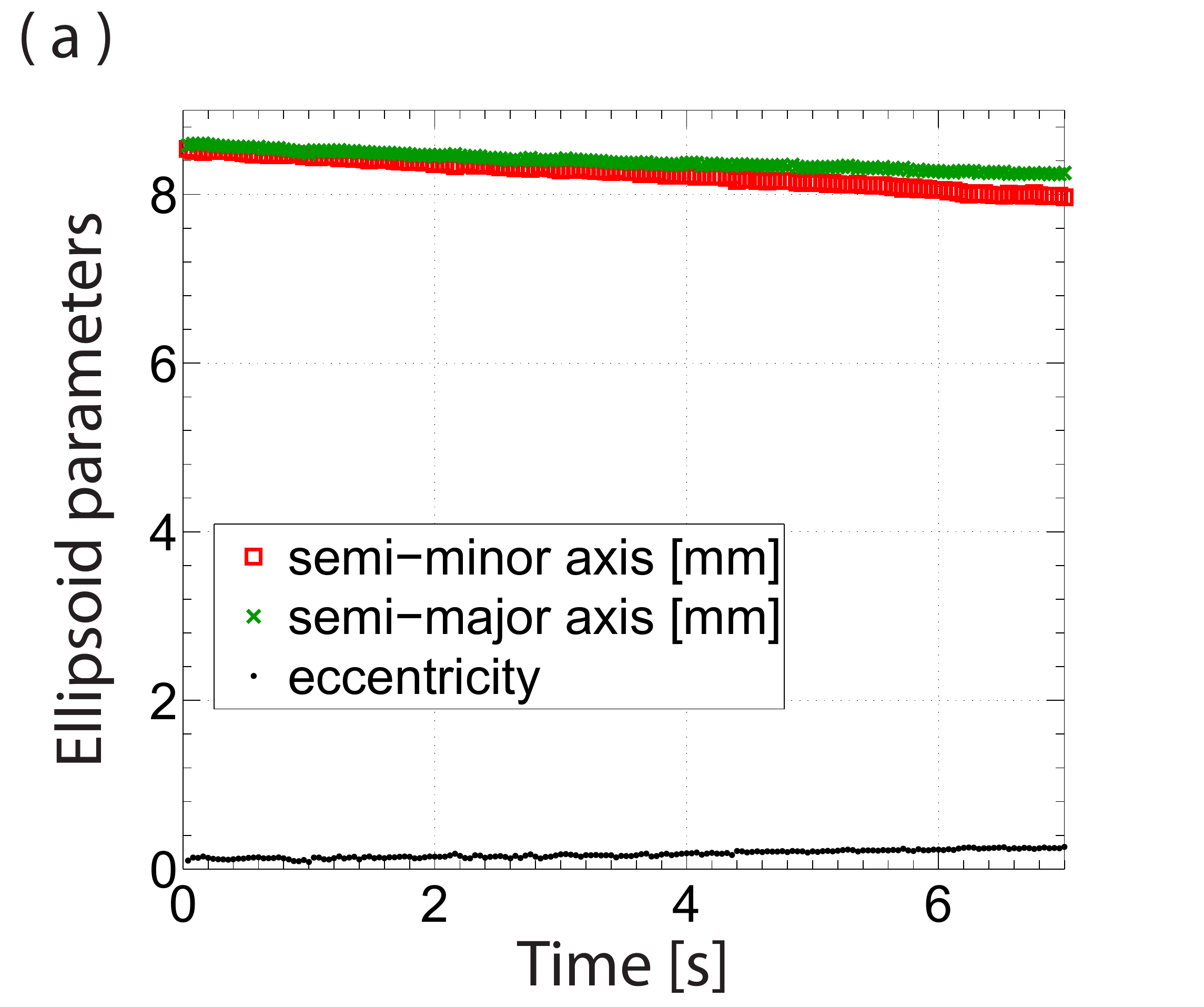} 
    \includegraphics[width=.49\columnwidth]{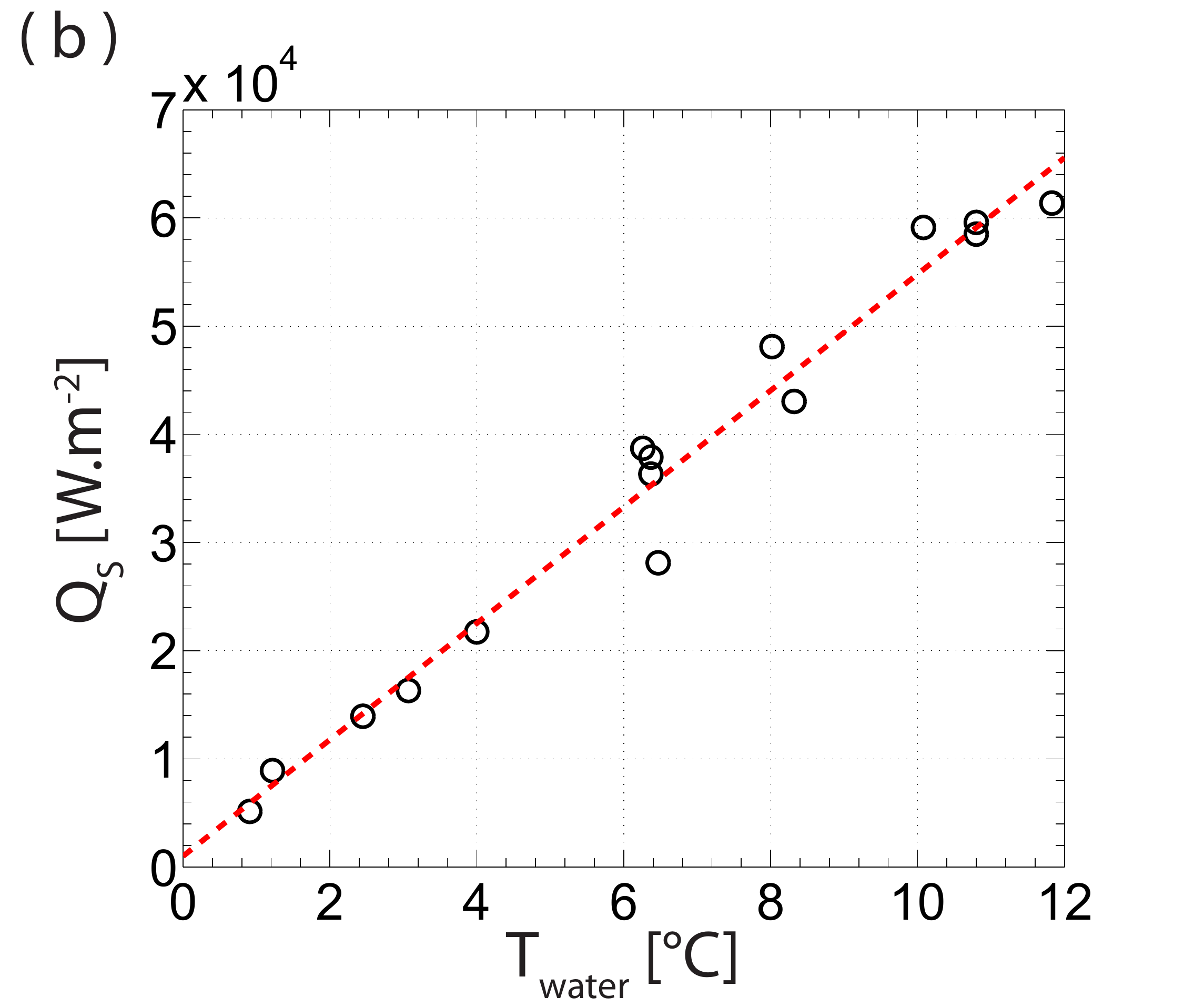}
    \caption{(colour online) (a): Evolution of the semi-minor and -major axes and eccentricity of a 18 mm ice ball fixed in the flow produced by two discs rotating at a frequency of 1.5 Hz. (b): Measure of the total heat flux for different flow temperatures with a constant rotation frequency and ice ball diameter. The red dotted line is a linear fit of expression: $Q_S=h(T_{water}-T_S)$ with $h=5380$ W.m$^{-2}$.K$^{-1}$ and $T_S=-0.2$°C.}
    \label{stefan}
  \end{center}
\end{figure}

We first illustrate the procedure with $18$ mm ice balls melting in a zero mean flow, maintained fixed at the centre of the 2 discs von K\'arm\'an flow. After the ice balls are made around the insulating rod and thermalized at melting temperature $0^\circ$ C, they are inserted into the flow at temperature $T_{water}$. We then record one movie per particle at a frame rate $F_{s}=25$ Hz and store all the movies on the computer for post processing. We finally measure the evolution of the particle shapes on the raw images using Matlab with image processing toolbox. In such an anisotropic turbulent flow, even if the ice balls remain nearly spherical in the first steps of the melting dynamics (several seconds), they slowly take an ellipsoidal (rugby ball) shape with major axis aligning with the axis of rotation $z$ of the experiment. This is illustrated on figure \ref{stefan}(a) where we have plotted the time evolution of the ellipsoid parameters (semi-minor axis $a$, semi-major axis $b$ and eccentricity $e$ defined in equation \ref{eq:surface}). We then restrict all the analysis to the $5$ first seconds of the movies for which the relative anisotropy of the particle is always less than $5\%$ (eccentricity less than $0.3$). For this short time interval the evolutions of $a$ and $b$ are linear and we estimate the volume as $V= {4} \pi a^2 b/{3}$, and the surface $S$ and eccentricity $e$ using the formulas:

\begin{equation}
S = 2 \pi a \left(1+b \frac{\arcsin(e)}{e} \right),  e=\frac{\sqrt{b^2-a^2}}{b}.
\label{eq:surface}
\end{equation}

By replacing the derivatives of $a$ and $b$ by the slopes obtained with linear fits of $a(t)$ and $b(t)$ in the five first seconds, we obtain a time-average value of $S^{-1}dV/dt$, noted $\overline{dR/dt}$.
In order to check the validity of equation (\ref{eq:stefan}) we have repeated this experiment for $18$ mm spheres at varying $T_{water}$ with fixed Reynolds number, and estimated the melting speed $\overline{dR/dt}$ averaged over the first five seconds of the experiments. As demonstrated in figure \ref{stefan}(b), the initial thermalization of the ice balls ensures the quantity $Q_{S}=\rho_{p}L_{f}\overline{dR/dt}$ to be of the form $Q_{S}=h (T_{water}-T_S)$ with good accuracy. The linear fit gives a surface temperature $T_S=-0.2^\circ$ C, very close to the melting temperature $T_0=0^\circ$ C, the difference being of the order of a possible offset of the thermometer. For a given rotation rate and ice ball diameter, finding a linear relation between $\Delta T$ and $Q_S$ also implies forced convection is indeed overwhelming natural convection in our problem. This is expected for such high Reynolds number flow with inertial steering and small imposed temperature differences. All scaling laws obtained in the following will thus only be based on forced convection arguments. For the next sections heat transfer coefficient $h=\rho_{p}L_{f}\overline{dR/dt}/(T_{water}-T_0)$ will be measured at varying particle sizes and rotation rates for only one flow temperature. From these measurements we will report evolution of the Nusselt number $\mbox{\textit{Nu}}=hD/\lambda_{th}$ quantifying the ratio between actual heat flux and diffusive heat flux estimated for a spherical particle of diameter $D$ with imposed temperature difference $\Delta T=T_{water}-T_0$.

\section{Results}

\subsection{Melting of fixed ice balls}

\begin{figure}[h]
  \begin{center}
    \includegraphics[width=.49\columnwidth]{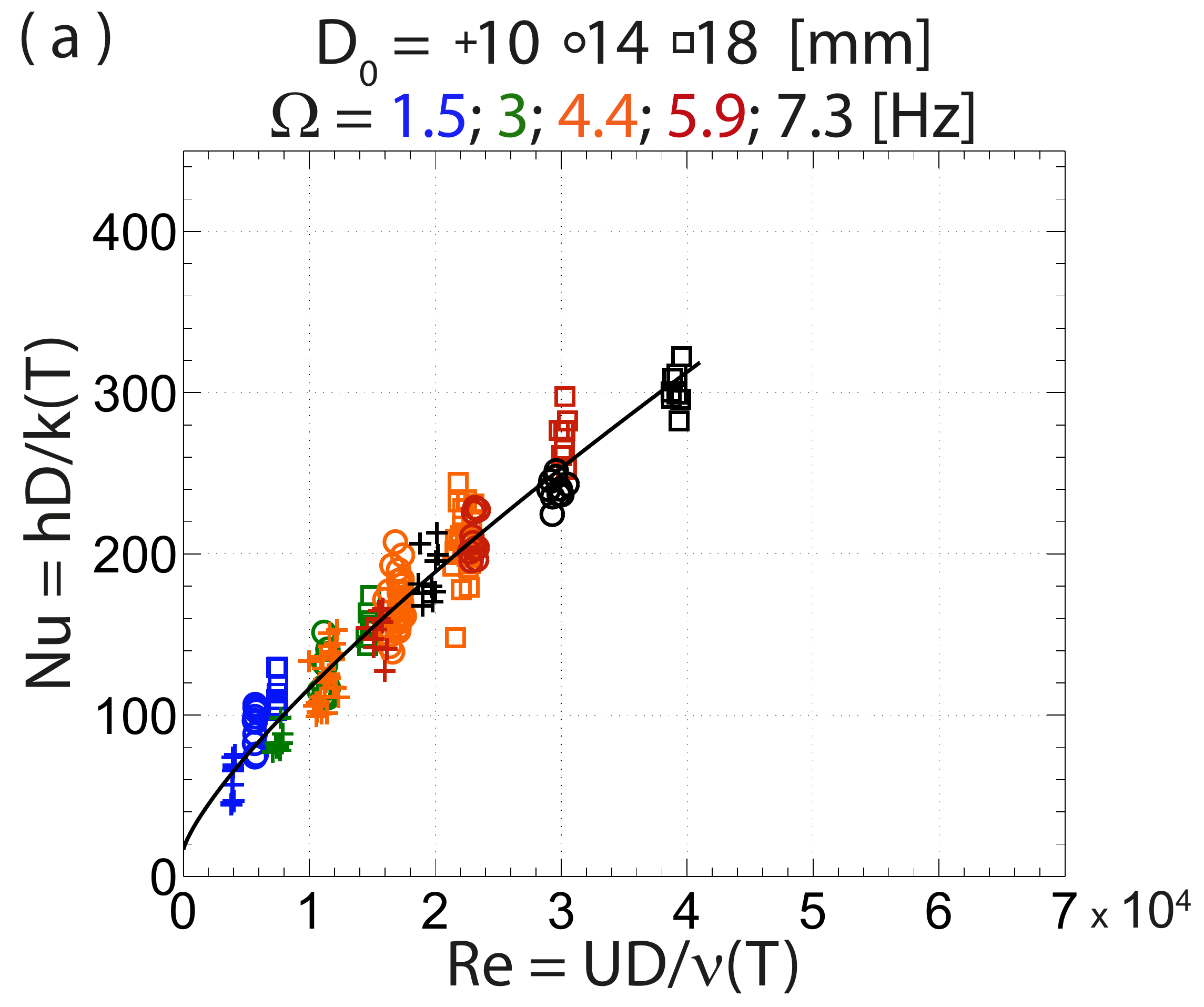}
    \includegraphics[width=.49\columnwidth]{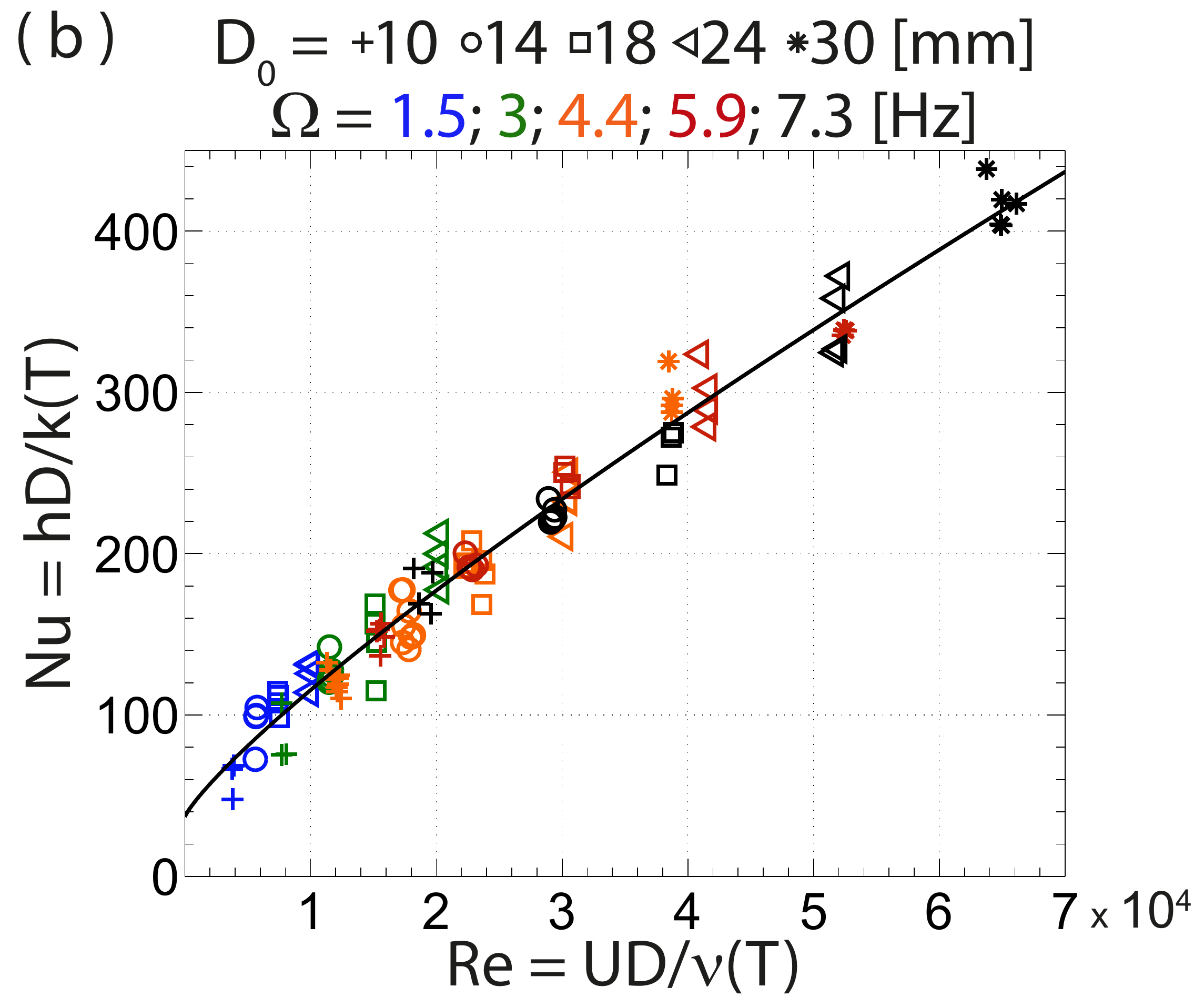}
    \caption{(colour online) Evolution of the particle Nusselt number as a function of the particle Reynolds  number $\mbox{\textit{Re}}_{D}=UD/\nu$ for ice balls fixed in the two discs (a) or one disc (b) flow. Different symbols are for different initial diameters $D$, different colours for different large scale velocities $U=2 \pi R \Omega$. The black lines are power law fits of expression $\mbox{\textit{Nu}}= \alpha+\beta\mbox{\textit{Re}}_{D}^{\gamma}$ with $[\alpha;~\beta;~\gamma]=[16~;7.5\cdot 10^{-2};~0.78]$ and $[37;~3.6\cdot 10^{-2};~0.84]$ for figure (a) and (b) respectively.}
    \label{Nu_fixe}
  \end{center}
\end{figure}

We first investigate the melting of ice balls maintained fixed in the turbulent flow and compare the two following situations: melting in a zero mean turbulent flow with $u'/U \sim 20\%$ (case A) and melting in a strong mean flow with $u'/U \sim 10\%$ (case C), both flows having the same large scale Reynolds number $\mbox{\textit{Re}}=UR/\nu$ with large scale velocity $U=2 \pi R \Omega$. This case is simpler than the melting of freely suspended particles (case B) because the slip velocity between the particle and the fluid may be estimated as the flow velocity measured in the absence of the ice ball.\\ 
The evolution of the measured Nusselt number as a function of the particles Reynolds number $\mbox{\textit{Re}}_{D}=UD/\nu$ is plotted in figure \ref{Nu_fixe}(a,b) for these two situations. In both cases we observe $\mbox{\textit{Nu}}$ to be in the range $[100,~400]$ for $\mbox{\textit{Re}}_{D}$ in the range $[5\cdot 10^3,~65\cdot 10^3]$. Although the turbulent flows are different, with very different values of time averaged and rms velocities, we find the Nusselt number to be of the same order of magnitude in both cases. This reveals the weak impact of the local turbulence level ($35\%$ or infinite around the particles respectively for the one disc and two discs flow) for heat or mass transfer in such fully turbulent flows. \\

For both configurations we find the Nusselt number to be a function of $\mbox{\textit{Re}}_{D}$ well fitted by an empirical power law $\mbox{\textit{Nu}} \sim \alpha + \beta \mbox{\textit{Re}}_{D}^\gamma$, as often reported in the literature \cite{vander,Levins1972,Birouk2002} (and references therein). The major difference between the two curves obtained is found to be in the scaling exponent $\gamma$, a quantity known to increase with increasing $\mbox{\textit{Re}}_{D}$. We find $\gamma=0.84$ for the one disc flow, of the same order but larger than the value $\gamma=0.78$ found for the two discs configuration. This may be due to the fact that computing the true rms values of the velocity for case A and C, one finds case C produces sliding velocities only $7-15\%$ larger than case A (where $\langle u\rangle=0$). The local based Reynolds number $\mbox{\textit{Re}}_{D}'=u_{trms}D/\nu$ are in the range $[1400,~23500]$ and $[1250,~20400]$ respectively for the one disc and two discs flow, which is consistent with the values found for $\gamma$. These values are much larger than the values reported for smaller particles dissolving in water \cite{Sano1974,Boon-Long1978} or $\gamma=2/3$ for evaporating droplets in air \cite{Birouk2002}, again consistent with the larger values of $\mbox{\textit{Re}}_{D}$ in the present experiments.

\subsection{Melting of freely advected ice balls}
\begin{figure}[h]
  \begin{center}
    \includegraphics[width=.49\columnwidth]{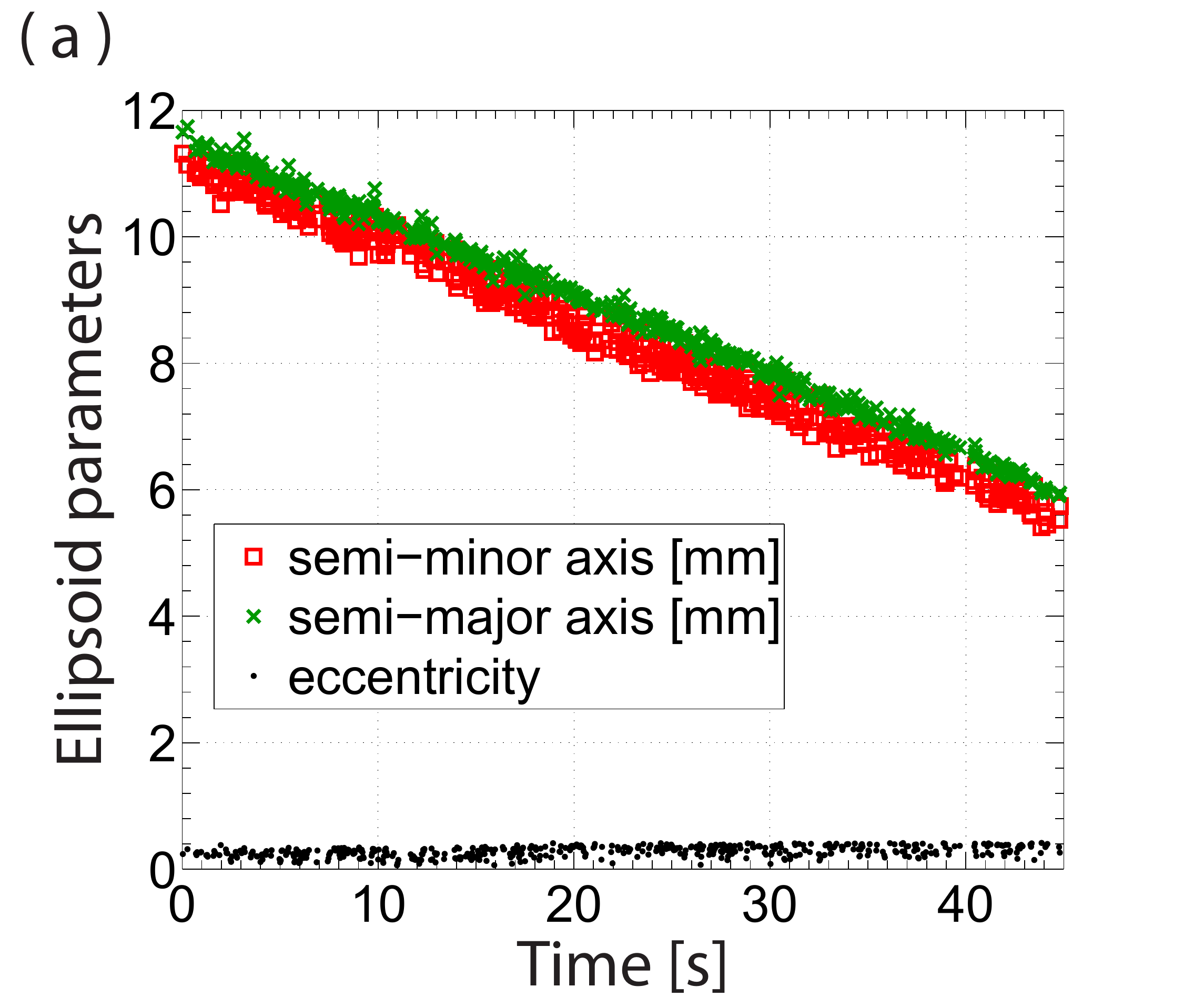}
    \includegraphics[width=.49\columnwidth]{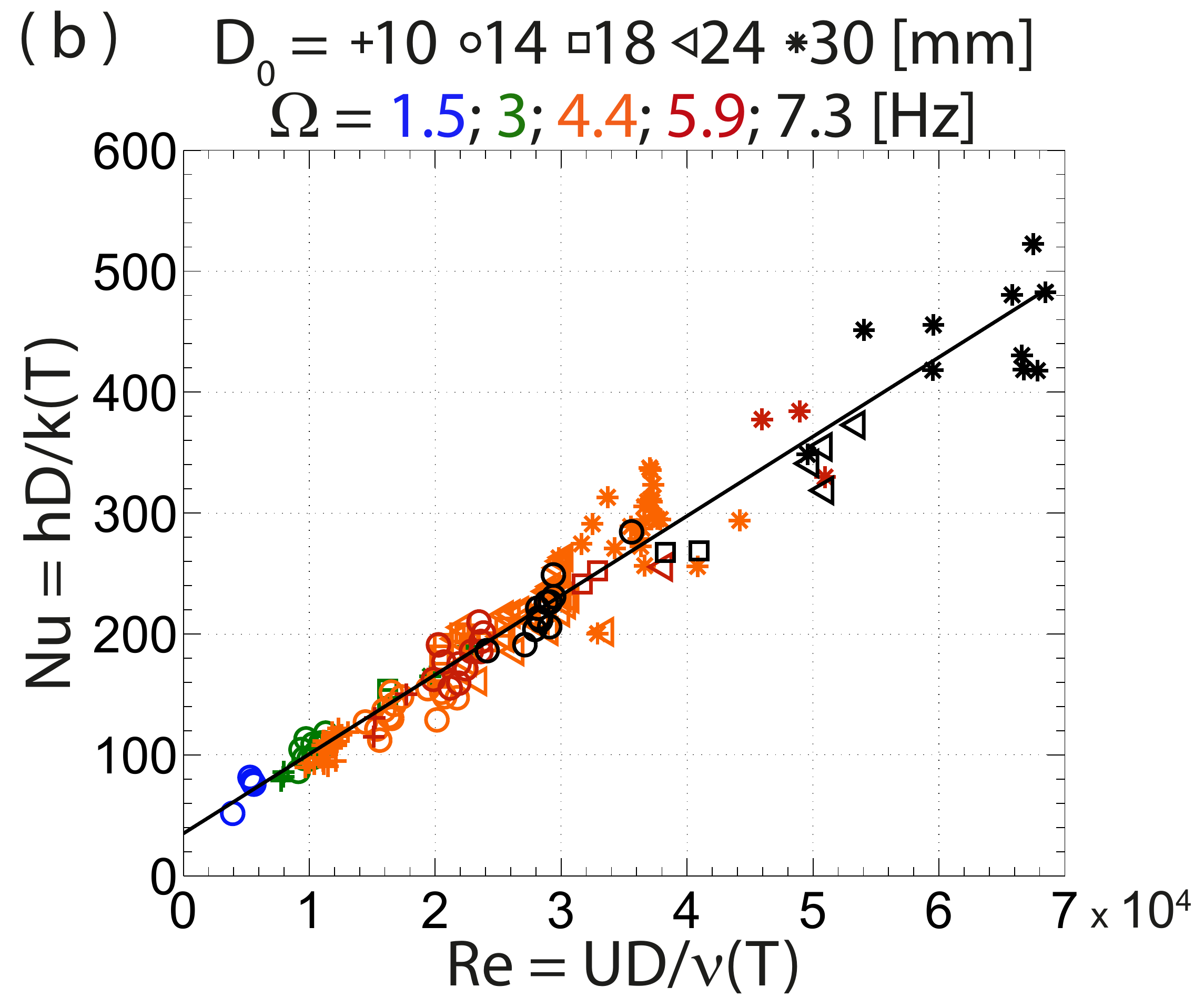}
    \caption{(colour online) (a): Evolution of the semi-minor and -major axes and eccentricity of a 24 mm ice ball advected by the flow produced by two discs rotating at a frequency of 4.4 Hz. (b): Evolution of the particle Nusselt number as a function of the particle Reynolds number for ice balls freely advected in the two discs flow. Different symbols are for different initial diameters $D$, different colours for different large scale velocities $U=2 \pi R \Omega$. The black line is a linear fit of expression $\mbox{\textit{Nu}}=\alpha+\beta\mbox{\textit{Re}}_{D}$ with $[\alpha;~\beta]=[35;~6.6\cdot 10^{-3}]$.}
    \label{Nu_libres}
  \end{center}
\end{figure}

We now turn to the case of freely advected particles melting in the 2 discs turbulent flow (case B). This case is more complex than the two previous situations because the heat transfer between the particle and the fluid depends on the sliding velocity, a quantity that depends not only on the flow characteristics, but also on the particle properties (size $D$, density $\rho_p$). The motion of particles with diameters of the order of the integral scale of the flow was only the topic of recent Lagrangian studies that revealed their translation and rotation dynamics are very intermittent as the particle explores the whole flow \cite{bib:zimmermann2011_PRL,Klein2013}; particles strongly modify the flow in their vicinity as compared to the situation when the particle is absent \cite{bib:naso2010_NJP,Klein2013}.
By following the moving particles in a large flow volume while measuring their shapes on raw images, it was possible to extend the analysis made in the case of fixed particles to the case of freely advected particles. We discovered that contrary to the fixed particle cases, for which the shape of the particle reflects the large scale anisotropy of the flow, the melting of freely suspended ice balls is isotropic, the particles remaining spherical for hundreds of large eddy turnover time $T=1/\Omega$. As demonstrated in figure \ref{Nu_libres}(a), the minor and major axes evolve in the same way, the little difference between them accounting for detection technique. They are indeed calculated by taking the smaller and bigger distances inside the detected object, hence small surface imperfections on a round object lead easily to an eccentricity around $0.2$, corresponding to a relative difference around $3\%$. 
The reason an ice ball remains spherical in such an anisotropic flow with inhomogeneous and anisotropic fluctuations is probably because of its rotation dynamics, which was proven to be coupled to its translation dynamics \cite{bib:zimmermann2011_PRL,Klein2013}. These studies also revealed large particles rms velocities is proportional to the large scale velocity $U=2 \pi R \Omega$ in von K\'arm\'an flows. In the absence of more information about the magnitude of the sliding velocity, we chose to display the measurements of Nusselt number as a function of the particle Reynolds number $\mbox{\textit{Re}}_{D}=UD/\nu$. 
Results are displayed in  figure \ref{Nu_libres}(b), which shows that the heat transfer magnitude of freely advected ice balls is comparable to the fixed particle cases, although $10\%$ larger. As opposed to the two previous cases, we now find the Nusselt number to be a linear function of the Reynolds number (figure \ref{Nu_libres}(b)). This result is very different from the correlations found for classical heat or mass transfer studies where exponents were always smaller than $0.85$ \cite{bib:birouk2006}. Our case corresponds to an ultimate regime of heat transfer for which the heat transfer coefficient $h$ is no longer dependent on the particle diameter $D$, but is only proportional to the rms value of velocity fluctuations.\\
This ultimate scaling of heat transfer is consistent with a fully turbulent hydrodynamical boundary layer around the particle, with a viscous sub-layer thickness $\delta_\nu$ much smaller than the diameter $D$. Indeed, for a fully developed turbulent boundary layer, one expects the wall shear stress to be of the order of $\tau^\star \sim \rho_{f} (v^\star)^2$, with $\rho_{f}$ the fluid density and $v^\star$ a skin friction velocity proportional to rms value of the fluid velocity $u'$. For such fully developed boundary layer, the viscous sub-layer is then of the order  $\delta_\nu \propto {\nu}/{u'}$, and is proportional to the inverse of the rms value of the velocity fluctuations. Following Reynolds analogy \cite{bib:tennekes_book}, the estimate of $\delta_\nu$ can be used as a measure of the thermal sub-layer thickness $\delta_T \sim \delta_\nu$ for fluids with Prantl number of order unity, which is the case for water. Finally one may estimate the heat flux per unit area $Q_{S}=\lambda_{th}\Delta T/\delta_{T}$ to obtain a linear relation between the Nusselt and Reynolds numbers $\mbox{\textit{Nu}} \propto \mbox{\textit{Re}}_{D}$. We note from this analysis the observed scaling law corresponds to the maximum exponent one may obtain from heat transfer measurements, and may be called the ultimate heat transfer regime of forced convection.

\section{Discussion and conclusion}
We have introduced a new measurement technique combining particle tracking and shadowgraphy, which allows for the sizing of moving ice balls with negligible variations in the apparent size with particle position, in nearly the whole volume of a turbulent von K\'arm\'an flow. From the evolution of size and shape of the particles we were able to measure the turbulent heat flux between the fluid and the ice balls as a function of the particle Reynolds number $\mbox{\textit{Re}}_{D}=UD/\nu$. With this measurement technique we studied the influence of turbulence on the heat transfer of melting ice balls in fully turbulent flows with high fluctuations. Three different cases were considered: freely advected ice balls, fixed ice balls under a mean drift much higher than the fluctuations and fixed ice balls under a mean drift much lower than the fluctuations. Varying the water temperature $T_{water}$ with all other parameters kept constant, we checked the relation between the surface flux $Q_{S}$ and the heat transfer coefficient $h$: $Q_{S}=h(T_{water}-T_0)$, confirming melting occurs close to thermal equilibrium.\\ 

For all cases, the Nusselt number was found to be very high and could be expressed as a power law of the Reynolds number. For the fixed particle cases, the exponents were found to be very high, close to $0.8$, with only weak impact of the turbulence level providing the true rms velocity are of the same order of magnitude. This is consistent with other studies \cite{bib:birouk2006} and might be expected for such fully turbulent flows because all velocities (mean and fluctuating) are proportional to the large scale forcing $U=2 \pi R \Omega$.
As opposed to fixed particle cases, freely advected ice balls were found to melt in an ultimate regime of heat transfer for which the Nusselt number is proportional to the Reynolds number. The result differs from what would have been expected from the remark that the sliding velocity for freely advected case should fall in between the cases of zero and large 
time average sliding velocities. The reason why the scaling law is different for freely advected particles is not presently known, the difference may come from the nature of the particle itself, which do not behave as a tracer of the flow motions, or from the fact that such large spherical particles rotate on themselves while moving, as was recently observed in similar turbulent flows \cite{bib:zimmermann2011_PRL,Klein2013}. This added degree of freedom might allow the hydrodynamic and thermal boundary layer to reach a fully developed regime on the particle surface leading to an ultimate regime of heat transfer, for which the heat flux per unit area no longer depends on the particle diameter $D$. Besides, the rotation dynamics has another important consequence. Although the shape of fixed particles was found to adapt rapidly to the anisotropic flow configuration, free particles were found to keep their spherical shape for hundreds of large eddy turnover times while exploring the whole flow volume. No matter the anisotropy and non-homogeneity of the flow turbulence, the ability of the particle to rotate on itself allows a conservation of its shape for very long times. This result may be useful for modellers who are interested in turbulent phase change of solid particles as it shows the possibility to model the particle as a non-deformable sphere. One may then try to compute heat transfer from large particles in practical configurations as a companion problem of particle laden flows (when simulated by an immersed boundary method) as was recently done for heavy particles transported in a channel flow \cite{kidanemariam:13}.

\begin{acknowledgements} This work is part of the International Collaboration for Turbulence Research, and was supported by ANR-12-BS09-0011 TEC2, and by labex IMUST from the Universit\'e de Lyon. We thank B. Castaing, M. Bourgoin, and N. Mordant for fruitful discussions, and LEGI Laboratory of Grenoble for sharing the PDI apparatus used for flow calibration.
\end{acknowledgements}
\bibliography{main}

\end{document}